\documentclass{aa}

\usepackage{graphicx}

\usepackage{ulem}
\usepackage{txfonts}

\begin{document}

\title{The effect of activity on stellar temperatures and radii}

\author{Juan Carlos Morales \inst{1}, 
        Ignasi Ribas \inst{1,2}
        \and  
        Carme Jordi \inst{1,3}}

\offprints{J.C. Morales}

\institute{Institut d'Estudis Espacials de Catalunya (IEEC), Edif.
Nexus, C/Gran Capit\`a, 2-4, 08034 Barcelona, Spain\\ 
\email{morales@ieec.uab.es}
\and
Institut de Ci\`encies de l'Espai (CSIC), Campus UAB, Facultat de Ci\`encies, Torre C5 - parell - 2a planta, 08193 Bellaterra, Spain\\
\email{iribas@ieec.uab.es}
\and
Departament d'Astronomia i Meteorologia, Universitat de Barcelona, Avda. Diagonal, 647, 08028 Barcelona, Spain\\
\email{carme@am.ub.es}
}
\date{Accepted November 13, 2007}
\authorrunning{Morales, Ribas \& Jordi}
\titlerunning{The effect of activity on temperatures and radii}

\abstract
{Recent analyses of low-mass eclipsing binary stars have unveiled a 
significant disagreement between the observations and the predictions of 
stellar structure models. Results show that theoretical models 
underestimate the radii and overestimate the effective temperatures of 
low-mass stars but yield luminosities that accord with observations. A 
hypothesis based upon the effects of stellar activity was put forward 
to explain the discrepancies.}
{In this paper we study the existence of the same trend in
single active stars and provide a consistent scenario to explain
systematic differences between active and inactive stars in the H-R 
diagram reported earlier.}
{The analysis is done using single field stars of spectral types late-K 
and M and computing their bolometric magnitudes and temperatures through 
infrared colours and spectral indices. The properties of the stars in 
samples of active and inactive stars are compared statistically to reveal 
systematic differences.}
{After accounting for a number of possible bias effects, active stars 
are shown to be cooler than inactive stars of similar luminosity therefore 
implying a larger radius as well, in proportions that are in excellent 
agreement with those found from eclipsing binaries.}
{The present results generalise the existence of strong radius and 
temperature dependences on stellar activity to the entire population of 
low-mass stars, regardless of their membership in close binary systems.}

\maketitle

\section{Introduction}

Structure and evolution of stars in the main sequence is, in principle, 
well understood. Theoretical models have made very significant progress in 
the past decades and are able to successfully fulfill most of the 
observational diagnostics (e.g., Lebreton 2000; Gallart et al. 2005). 
However, there are still a few open issues that taint the optimistic 
picture. Most notably, the theoretical modeling of the structure and 
evolution of stars on both ends of the main sequence is not yet fully 
resolved (e.g., Ribas 2006a).

Model testing in the low-mass regime has been usually restricted to the 
evaluation of the mass-luminosity relationship because of the relatively 
large availability of observational data. Studies such as Delfosse et al. 
(2000) concluded that models agree well with empirical measurements, 
especially when the tests are performed in near-infrared bands, which are 
less affected by metallicity (Chabrier \& Baraffe 2000; Bonfils et al. 
2005). The detailed study of the M-type eclipsing binary YY Gem by Torres 
\& Ribas (2002) contributed to the tests by providing simultaneous and 
accurate measurements of the component masses, radii, effective 
temperatures and luminosities. The surprising result was that the 
components of the system were found to be significantly larger and cooler 
than the predictions of stellar structure models, but yet with 
luminosities in agreement with theoretical calculations. The same trend 
was later confirmed by a number of subsequent analyses of other eclipsing 
binaries (see Ribas 2006b and L\'opez-Morales 2007 for a complete 
summary). Thus, a growing body of observational evidence is challenging 
the adequacy of current stellar models in reproducing the overall 
properties of low-mass stars.

Using the available data and analysing the possible scenarios, Ribas 
(2006b) and Torres et al. (2006) concluded that a plausible explanation 
for the observed discrepancies could be related to stellar activity. In 
this respect, the active components of eclipsing binaries appear to be 
larger and cooler than inactive single stars (i.e., those reproduced by 
theoretical models) while keeping similar luminosity. To first 
approximation, this means that, regardless of changes in the stellar outer 
layers, the rate of nuclear burning in the core is not modified by 
activity and therefore the overall flux is conserved. The mechanism 
responsible for the observed radius and temperature differences has not 
been identified yet, although the efficiency of convection in strong 
magnetic fields (Mullan \& MacDonald 2001), or the simple effect of flux 
conservation in a spot-covered stellar surface (L\'opez-Morales \& Ribas 
2005) could explain the observed changes in the stellar properties. Note 
that a possible correlation of the radius differences with stellar 
metallicity could also be present (Berger et al. 2006).

The stellar activity hypothesis has been built in the context of close 
binary systems, where orbital synchronization forces the components to 
rotate fast, therefore triggering high levels of magnetic activity. With a 
much longer history, the question of the possible differences between 
active and inactive stars has been a recurrent one (Kuiper 1942; Joy \& 
Abt 1974). Recent studies have been generally focused on the comparison of 
radiative properties such as spectral types or photometric colours. The 
conclusions have been quite diverse, both in favour (Hawley et al. 1996; 
Amado \& Byrne 1997) and against (West et al. 2004; Bochanski et al. 2005) 
the existence of systematic colour differences. One of the most conclusive 
analyses is that of Stauffer \& Hartmann (1986), where the authors 
identified a separation of the sequences of active and inactive stars in a 
luminosity-colour plot. The reasons for the observed separation in the 
sequences could not be unambiguously identified since it could both come 
from luminosity differences at constant effective temperature, effective 
temperature differences at constant luminosity, or a combination of the 
two.

The results from eclipsing binaries provide a new perspective to analyse 
this long-standing issue since temperatures, radii and luminosities can be 
determined in a fundamental manner. As discussed above, the components of 
eclipsing binaries seem to be cooler and larger than model predictions but 
with luminosities in agreement. Very importantly, this observation is 
confirmed by the theoretical study of Chabrier et al. (2007), who have 
recently modified stellar evolution codes to include the effects of 
stellar activity. These authors also report systematic differences 
between the properties of active and inactive stars in similar amount and 
trend as those found from eclipsing binary studies.

With improved statistics with respect to Stauffer \& Hartmann (1986) and 
the context provided by the new evidence discussed above, we analyse the 
existence of differences between active and inactive stars of spectral 
types late-K and M. In the present study we use luminosities directly 
determined from accurate trigonometric parallaxes and carry out a thorough 
analysis of possible biases, such as the effect of pre-main sequence (PMS) 
and binary stars. Also, we interpret the results in physical terms, i.e., 
effective temperature and radius variations. If active stars were indeed 
cooler and larger than their inactive counterparts, while keeping similar 
luminosities, this should be observable in single field stars (in addition 
to close binaries) thus generalizing the proposed stellar activity 
scenario to all low-mass stars.

\section{Sample of late-K and M stars}

The sample used to test this hypothesis is composed of selected late-K and 
M dwarfs from the Palomar/Michigan State University survey of nearby stars 
(hereafter PMSU; Reid et al. 1995; Hawley et al. 1996). This catalog lists 
the position, $M_{\rm V}$, distance, TiO, CaH and CaOH spectral indices, 
H$\alpha$ equivalent width and proper motions for each of the 1966 stars 
that it contains. The distances listed are averaged combinations 
of Hipparcos trigonometric parallaxes and spectrophotometric 
determinations. Because of our working hypothesis of the $T_{\rm eff}$ 
dependence on activity, only objects with direct trigonometric parallaxes 
are useful to our study because spectroscopic and photometric parallaxes 
could be biased by activity. The restriction of trigonometric parallaxes 
reduces the number of stars in the working sample to 746, with 
1.3~$<d<$~58.0~pc and 6.65~$<M_{\rm V}<$~16.0 mag.

To calculate the bolometric magnitude (i.e., luminosity) of the sample 
stars we made use of photometry in the IR bands because of the weaker 
dependence on the bolometric correction ($BC$), which could be a potential 
source of large uncertainty (for M-type stars variations of 200~K in 
$T_{\rm eff}$ produce only changes below 0.1 mag in $BC_{\rm K}$). Thus, 
the comparison between active and inactive stars with the same luminosity 
is more reliable if $M_{\rm bol}$ is computed from the $K$ band rather 
than the $V$ band. The former is also less affected by variability caused 
by surface inhomogeneities. The sample was cross-matched with the 2MASS 
and ROSAT survey catalogs to obtain $J$, $H$ and $K_{\rm s}$ 
magnitudes and $L_{\rm X}$ for each star. $K_s$ was 
transformed to the $K$ Johnson band (Alonso et al. 1994) to compute 
$BC_{\rm K}$ using the models in Bessell et al. (1998) as a function of 
$T_{\rm eff}$. The available trigonometric distances were used to compute 
$M_{\rm K}$ and, subsequently, $M_{\rm bol}$. The TiO5 index was used as 
spectral type indicator with the prescription on Reid et al. (1995), which 
suggests a linear relationship with small scatter between this 
spectral index and the spectral subtype of M stars
($\mbox{Sp.Typ.}=-10.775\cdot \mbox{TiO5}+8.2$). Effective temperatures 
were derived using the spectral type-temperature correspondence in Bessell 
(1991) -- which is similar to that of Leggett et al. (1996) -- and the 
equivalent width of the H$\alpha$ emission line was used as an indicator 
of magnetic activity. An iterative procedure was used to ensure 
consistency between the adopted $T_{\rm eff}$ and $BC_{\rm K}$.

\section{Discussion}

Active and inactive stars were considered separately in the analysis of 
the sample. Those stars with the H$\alpha$ line in absorption were 
classified as inactive while those with H$\alpha$ in emission were 
considered to be active. Note that this criterion only identifies as 
active stars those with high levels of activity, since mildly active stars 
can still have H$\alpha$ in absorption (e.g., Cram \& Mullan, 1979). 
Following this approach, 72 stars out of the total 746 turn out 
to be active. In Tables 1 and 2, available electronically from CDS, we 
provide the lists of active and inactive stars, respectively, together 
with the parameters relevant to the present study.

It is important to emphasise that our analysis is only meaningful if both 
sets of active and inactive single stars are equivalent in terms of 
evolution and metallicity. In the case of late-K and M-type stars this 
means that the calculations need to be restricted to main sequence stars. 
The relatively short PMS evolutionary phase, in which stars are also 
magnetically active, would break the one-to-one correspondence between 
luminosity and mass (see below) and therefore invalidate the comparisons. 
Thus, we cleaned the active star sample from possible PMS objects. There 
are only a few known young star associations and moving groups in the 
solar neighbourhood and these have well-defined space motions. Using the 
catalogs of young moving groups (L\'opez-Santiago et al. 2006; D. 
Fern\'andez, priv. comm.), we removed a total of 22 stars (9 of 
them classified as active) from the sample that belong to ensembles with 
ages younger than $\sim$200 Myr. Since this is a crude approach, we 
carried out an independent theoretical estimation of the expected number 
of PMS stars as a function of mass for comparison. Using the models of 
Baraffe et al. (1998) and assuming a constant star formation rate in the 
solar neighbourhood, we estimated the fraction of PMS stars just by 
calculating the ratio of the time spent in the PMS phase and the total 
time during which we would classify the star as active. This calculation 
yielded a fraction of PMS stars of 20\% to 10\% (decreasing with mass) for 
our sample parameters. Such values turn out to be close to the actual 
fractions found when considering kinematic restrictions. For similar 
reasons, low-metallicity halo stars could also alter the results of the 
inactive star bins. We removed high-velocity stars and also subdwarfs 
using the CaH2-TiO5 prescription given in Bochanski et al. (2005). A total 
of 7 stars were eliminated at this step. In this way, both the inactive 
and active star samples contain disk stars and should have largely similar 
metallicities. Finally, we removed close binary stars from the list since 
we aimed to test if single active stars show the same discrepancies with 
the models as close binaries. We identified and rejected a total of 22
 binaries (14 with H$\alpha$ in emission) using SIMBAD and the 
lists given in Gizis et al. (2002) for the PMSU survey. The last columns 
of Tables 1 and 2 include detailed information on the stars that have been 
rejected using these criteria.

\begin{figure}[t]
\centering
\includegraphics[width=\columnwidth]{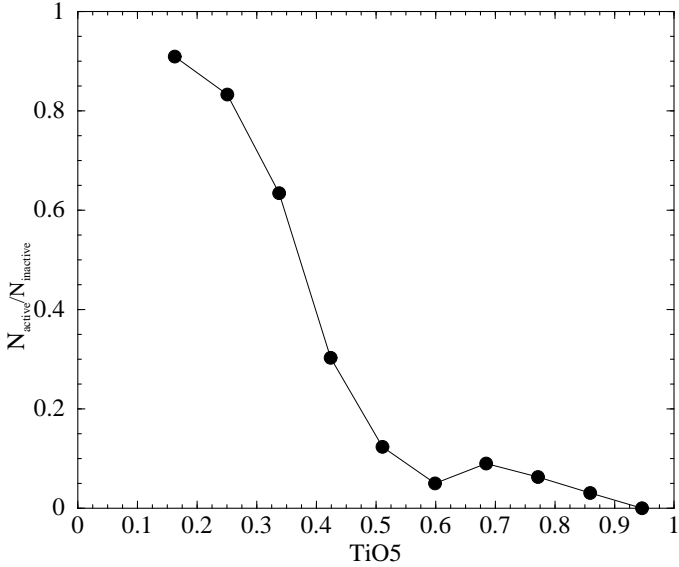}
\caption{Ratio of active to inactive stars as function of the TiO5 index
for the full PMSU sample. Active stars are defined as those with H$\alpha$
emission.}
\label{fig_cat0}
\end{figure}

We grouped the stars in $M_{\rm bol}$ (i.e., luminosity) bins. The bin 
size was selected to be 1 magnitude to keep a statistically significant 
number of stars in each bin. Note that, since the mass-luminosity 
relationship is so tight for low-mass stars (because post-ZAMS evolution 
is so slow), mass and luminosity bins are equivalent. Because of the need 
to use fairly large $M_{\rm bol}$ bins, our results could be potentially 
affected by some biases and selection effects. Firstly, a selection effect 
could be caused by the use of H$\alpha$ emission as activity criterion, 
which is more sensitive for stars of lower photospheric luminosities. 
Secondly, it is well known that the number of active stars increases 
towards later spectral types, i.e., decreasing effective temperature 
(e.g., West et al. 2004), which could tend to give more weight to the 
cooler active stars in each luminosity bin. Fortunately, both effects can 
be corrected using the statistics available in the PMSU sample itself. We 
have calculated the ratio of active to inactive stars as a function of the 
TiO5 index, as shown in Fig. \ref{fig_cat0}, in which active stars are 
selected using the H$\alpha$ emission criterion. To correct for the 
biases, the parameters related to active stars are weighted with the 
quantity $N_{\rm inactive}/N_{\rm active}$ estimated at their respective 
TiO5 index values in all subsequent calculations. The correction has a 
small -- albeit non negligible -- effect on the mean TiO5 indexes of the 
bins. As a check, we have calculated the same corrections using the active 
star ratios in West et al. (2004) and Bochanski et al. (2005) and the 
results are identical.

\begin{figure}[t]
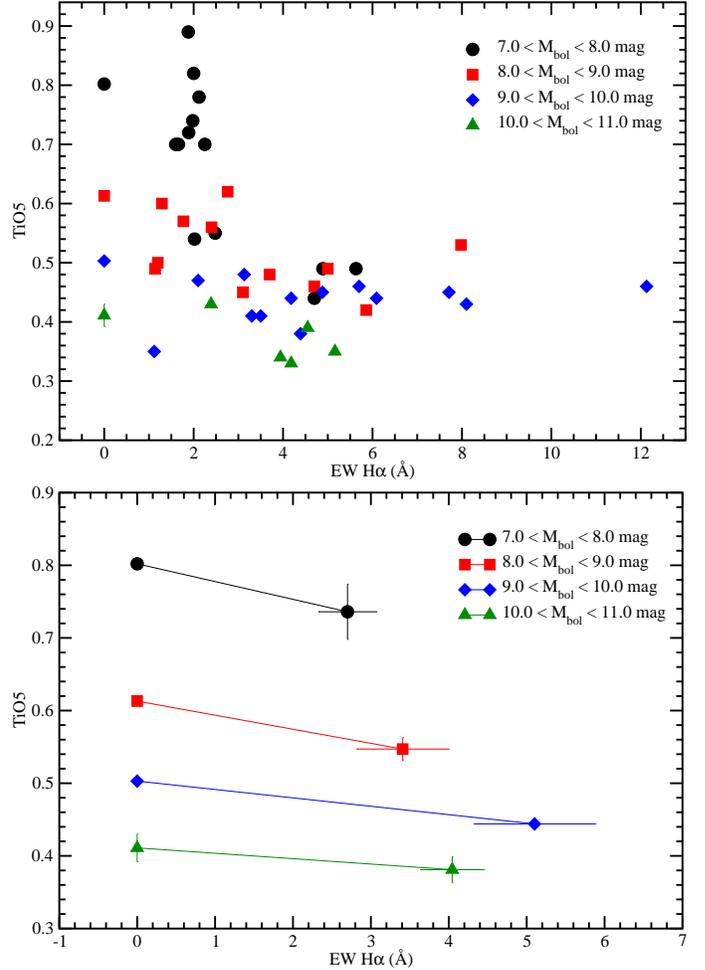

\centering
\includegraphics[width=\columnwidth]{fig2a.eps}
\includegraphics[width=\columnwidth]{fig2b.eps}
\caption{Distribution of active and inactive stars in TiO5 vs. EW 
H$\alpha$ for bins with statistical significance. Top: Single values for 
each active star and mean values for inactive stars. Bottom: Mean values 
for each $M_{\rm bol}$ bin.}
\label{fig_cat1}
\end{figure}

Figure \ref{fig_cat1} illustrates the distribution of stars in the TiO5 
versus H$\alpha$ equivalent width (EW) diagram (H$\alpha$ EW = 0 means 
line in absorption) for the 695 single main sequence, disk stars (48 
active and 647 inactive). $M_{\rm bol}$ bins with few stars 
(typically 2 or less) were not considered in the figure and subsequent 
comparisons and have not been reported in Table \ref{tab_means}. 
The average TiO5 index for each $M_{\rm bol}$ bin is shown 
in Fig. \ref{fig_cat1} for active and inactive stars separately. In Table 
\ref{tab_means} the differences between active and inactive stars are 
listed for statistically significant bins. Also provided are the 
average values of log($L_{\rm X}/L_{\rm bol}$) for active stars, which are 
at the saturation level (e.g., Pizzolato et al. 2003) for all luminosity 
bins. This indicates that our active sample is representative of stars 
with very high activity levels. It is obvious both from the plot and from 
the table that active stars have systematically lower TiO5 indices (i.e., 
lower effective temperatures) than their inactive counterparts of similar 
luminosity.

\addtocounter{table}{2}
\begin{table*}[t]
\centering
\caption[]{Differences between mean values for active and inactive 
stars for each bin of $M_{\rm bol}$ with statistical significance.} 
\label{tab_means}
  \begin{tabular}{c c c c c c c c}
   \hline\hline
   $M_{\rm bol}$ bin & $N_{\rm inactive}$ & $N_{\rm active}$ & $<\Delta TiO5>$ & $<\Delta T_{\rm eff}>$ (K) & $<\Delta R/R>$ (\%) & $<\Delta (V-K)>$ & $<$log($L_{\rm X}/L_{\rm bol}$)$>_{\rm active}$\\
   \hline
   7.0 -- 8.0   & 286 & 13 & $-$0.066$\pm$0.038 & $-$128$\pm$62 &  6.9$\pm$3.5 & 0.34$\pm$0.26 & $-$3.11$\pm$0.03   \\
   8.0 -- 9.0   & 208 & 12 & $-$0.066$\pm$0.018 & $-$107$\pm$29 &  6.3$\pm$1.8 & 0.31$\pm$0.07 & $-$3.19$\pm$0.11   \\
   9.0 -- 10.0  & 72  & 13 & $-$0.059$\pm$0.012 & $-$118$\pm$22 &  7.3$\pm$1.4 & 0.31$\pm$0.08 & $-$2.87$\pm$0.14   \\
   10.0 -- 11.0 & 13  & 5  & $-$0.030$\pm$0.026 &  $-$59$\pm$50 &  3.8$\pm$3.3 & 0.30$\pm$0.24 & $-$3.30$\pm$0.11   \\
   \hline
  \end{tabular}
\end{table*}

The mean TiO5 differences in Table \ref{tab_means} are all positive and 
quite similar in all magnitude bins. These index differences can be 
transformed into temperature differences using the calibrations described 
above and eventually into radius differences just by assuming a fixed 
luminosity ($L_{\mathrm{active}}\simeq L_{\mathrm{inactive}}$): 
\begin{equation}
 \frac{R_{\mathrm{ac}}-R_{\mathrm{inac}}}{R_{\mathrm{inac}}}=\left( 
             \frac{T_{\mathrm{eff,inac}}}{T_{\mathrm{eff,ac}}} 
           \right) ^{2}-1 
\end{equation}
These values are also listed in Table \ref{tab_means}. 

\begin{figure}[t]
\centering
\includegraphics[width=\columnwidth]{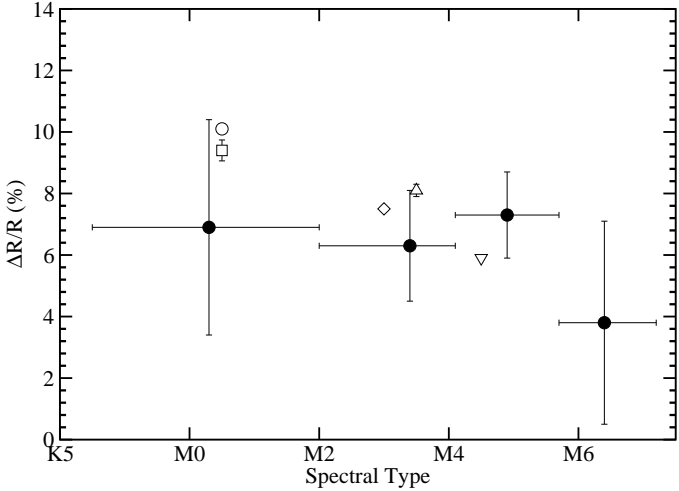}
\caption{Comparison of the differentials of radii over spectral type (filled 
circles) with results from the eclipsing binaries with highest accuracy (open 
symbols): YY Gem (circle), GU Boo (square), NSVS01031772 (diamond), CU Cnc (up 
triangle) and CM Dra (down triangle).}
\label{fig_cat2}
\end{figure}

Since differences in radius are computed through temperature ratios and we 
are carrying out a differential study, the results are almost independent 
of the $T_{\rm eff}$ scale adopted, which is still controversial for 
M-type stars. A cross-check with the $T_{\rm eff}$ calibrations of 
Schmidt-Kaler (1982) and de Jager \& Nieuwenhuijzen (1987) 
expectedly yielded the same results. The trend of lower temperatures and 
larger radii for active stars has high statistical significance since 
similar differentials are obtained in all luminosity bins. In Fig. 
\ref{fig_cat2} we show the calculated radius differences translated into 
spectral type bins and together with the radius differences obtained from 
a sample of eclipsing binaries with data of best quality (taken from 
L\'opez-Morales et al. 2006 and L\'opez-Morales 2007). Note that the two 
radius differentials have slightly different meanings. While eclipsing 
binary values come from the direct comparison of radius measurements with 
the predictions of theoretical models of Baraffe et al. (1998) -- which do 
not include the effects of stellar activity, -- the values for single 
stars are computed from the difference between active and inactive 
samples. As can be seen, the differentials from these two completely 
independent approaches are in very good agreement.

\begin{figure}[t]
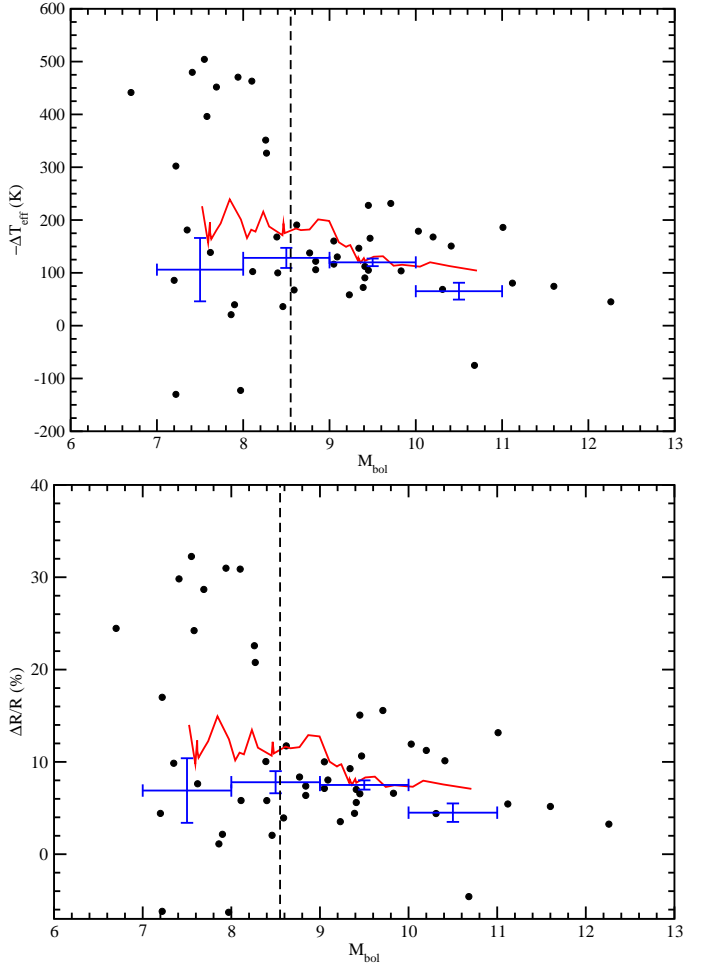

\centering
\includegraphics[width=\columnwidth]{fig4a.eps}
\includegraphics[width=\columnwidth]{fig4b.eps}
\caption{Differentials of $T_{\rm eff}$ (top) and $R$ (bottom) $vs.$ $M_{\rm
bol}$ computed from a polynomial fit to the $T_{\rm eff}$ of inactive stars.
Solid lines are 10-point running averages and the error bars show
the averages for four relevant $M_{\rm bol}$ intervals. The dashed line
marks the limit between fully convective stars and those with a radiative core.}
\label{fig_cat4}
\end{figure}

To rule out a possible effect of the $M_{\rm bol}$ binning procedure 
itself on the differences reported, we used an alternative approach using 
a polynomial fit. We first calculated a best-fitting third order 
polynomial of the form $T_{\rm eff}=f(M_{\rm bol})$ for the inactive star 
sample. Then, for each individual active star we calculated the 
temperature difference (and radius difference) between the observed value 
and the one predicted by the polynomial fit. A graphical representation of 
the results is provided in Fig. \ref{fig_cat4}, where the individual 
temperatures and radius differences are shown together with a running 
average of 10 points and averages computed for each $M_{\rm bol}$ bin. The 
actual values are given in Table \ref{tab_polynomial}, and are very 
similar to those in Table \ref{tab_means}, which were calculated from 
$M_{\rm bol}$ bins. An interesting feature of Fig. \ref{fig_cat4} is the 
relatively large scatter at $M_{\rm bol}<8.5$ mag, which occurs near the 
value where models predict the change between fully convective stars and 
those with a radiative core.

\begin{table}[t]
\centering
\caption{Effective temperature and radius differences of active stars
calculated from a polynomial fit to inactive stars.}
 \label{tab_polynomial}
  \begin{tabular}{c c c c}
    \hline\hline 
    $M_{\rm bol}$ bin & $N_{\rm active}$ & $<\Delta T_{\rm eff}>$ (K) & $<\Delta R/R>$ (\%) \\
    \hline
    7.0 -- 8.0   & 13 & $-$106$\pm$60 & 6.9$\pm$3.5    \\
    8.0 -- 9.0   & 12 & $-$128$\pm$19 & 7.8$\pm$1.2    \\
    9.0 -- 10.0  & 13 & $-$120$\pm$7  & 7.5$\pm$0.5    \\
    10.0 -- 11.0 &  5 &  $-$65$\pm$16 & 4.5$\pm$1.0    \\
    \hline
    \end{tabular}
\end{table}

A further useful check of the results comes from restricting the 
analysis to only stars that have been explicitly classified as single 
through high-resolution spectroscopy (Gizis et al. 2002). Although the 
statistics are less significant (with 127 inactive and 19 active stars), 
the mean temperature and radius differences for each $M_{\rm bol}$ bin 
are within one sigma of those in Table \ref{tab_means}.

In Table \ref{tab_means} we also include the mean differences in the 
$(V-K)$ colour index, which are a direct consequence of the different 
average $T_{\rm eff}$ values between active and inactive stars. The 
relatively large scatter of these means (especially that of the first 
$M_{\rm bol}$ bin) may be caused by the variability in the $V$ band of the 
stars in the active sample. This stems from the existence of 
surface spots with various cycles that could affect single-epoch $V$-band 
measurements but not $K$-band measurements that are more immune to 
spot-induced variability. Note that no obvious or only marginal colour 
differences for active stars have been reported before (see, e.g., Hawley 
et al. 1996; Bochanski et al. 2007). This would seem to stand in 
contradiction with our results, but it is not. It is important to 
emphasise that such colour comparisons are always carried out using 
spectral types as fiducials. Thus, the lack of colour differences between 
active and inactive stars of the same spectral types is just revealing 
that they have the same temperatures and spectral energy distributions, 
which is not surprising. The colour differences we find correspond to 
active and inactive stars of the {\em same luminosity}, which is 
equivalent to mass, and therefore a fundamental stellar property. Further, 
these are in good agreement with the differences in the spectral 
energy distribution caused by activity found by Stauffer et al. (2003) in 
Pleiades K-type stars.

A possible effect that could alter the outcome of this study is the 
influence of metallicity. Obviously, a metal-poor star will have 
intrinsically weaker TiO bands and therefore appear as hotter when the 
TiO5 index is used. For example, a metallicity decrease of 0.5 dex 
corresponds to a $\sim200$~K higher effective temperature. If there was a 
mean metallicity difference between the two samples used (active and 
inactive) this could affect their mean temperature difference (and 
inferred radius difference) based on the TiO5 index. There is certainly a 
metallicity spread within the two samples used, characteristic of a disk 
population, but the key issue here is whether the two samples have similar 
mean metallicities. A possible approach to test this is by directly using 
the metallicity calibration of Bonfils et al. (2005). However, this 
calibration is based in the $V-K$ index and, according to our hypothesis 
of an activity effect, the systematic differences in this index will cause 
a spurious systematic metallicity difference.

It is reasonable to assume that, on average, the inactive sample will be 
older than the active sample, just because there is a well-established 
age-activity relationship (e.g., Skumanich 1972). However, this age 
difference does not imply a difference in the mean metallicities of both 
samples because numerous studies have concluded that there is no 
age-metallicity correlation in disk stars, as shown by Nordstr\"om et al. 
(2004) and references therein. The average metallicity of the Nordstr\"om 
et al. catalog, when applying the kinematic constraints of our 
sample (i.e., $-165<U<130 $~km~s$^{-1}$, $-130<V<40 
$~km~s$^{-1}$ and $-90<W<80 $~km~s$^{-1}$), is $[M/H]=-0.17$. If we 
use the Bonfils et al. (2005) calibration in our inactive sample, whose 
stars should have unbiased $(V-K)$ indices, we obtain $[Fe/H]=-0.14$. Both 
averages are in good agreement and indicate that our inactive sample is 
representative of the overall population of the solar neighbourhood. If we 
apply the Bonfils et al. calibration to the active sample, also under the 
assumption that their $(V-K)$ indices are unbiased, we find a mean 
metallicity that is about 0.3 dex higher (i.e., $[Fe/H]\sim+0.15$). Such 
high mean metallicity value is very unlikely in the context of the 
Nordstr\"om et al. results, thus suggesting that the $(V-K)$ indices of 
the active star sample are indeed biased because of the effect of 
activity.

We carried out additional tests to eliminate the possibility of 
metallicity effects in our results arising from contamination by 
metal-poor stars belonging to the thin disk. When selecting only stars 
with $UV$ values in the interval ($-$90,50) km~s$^{-1}$, which would be 
characteristic of the overall thin disk, the results are nearly identical 
to those in Table \ref{tab_means}. We even considered a further test by 
sacrificing statistical significance but adopting very restrictive 
kinematic criteria in both the active and inactive star samples to ensure 
that both belong to the young disk population. If metallicity was 
responsible for the observed differences in the radii and temperature, we 
should expect them to disappear when both samples come from the same 
population. Thus, we selected stars belonging to the young disk using the 
$UV$ criteria of Montes et al. (2001) plus $-25 < W < 25$ km~s$^{-1}$. The 
resulting radius differences are listed in Table \ref{tab_difFeH}. As can 
be seen, the differences are fully compatible within the error bars with 
those of full sample, therefore indicating that metallicity effects do not 
play a significant role in our conclusions.

\begin{table}[t]
\centering
\caption{Active/inactive star radius differences using restrictive kinematic
criteria (see text for further details).}
 \label{tab_difFeH}
  \begin{tabular}{c c c}
    \hline\hline 
    $M_{\rm bol}$ bin & $N_{\rm inactive}/N_{\rm active}$ & $<\Delta R/R>$ (\%) \\
    \hline
    7.0 -- 8.0   & 68/8 & 11.6$\pm$4.2    \\
    8.0 -- 9.0   & 39/3 &  9.1$\pm$2.9    \\
    9.0 -- 10.0  & 19/5 &  4.9$\pm$2.3    \\
    10.0 -- 11.0 & 1/3$^{\mathrm{1}}$     \\
    \hline
    \end{tabular}
 \begin{list}{}{}
  \item[$^{\mathrm{1}}$] Mean not computed because of insufficient statistics.
 \end{list}
\end{table}

Stars with direct radius measurements will provide the ultimate proof of 
the systematic radius deviations with activity. The first steps in this 
direction have already been taken by L\'opez-Morales (2007), who used 
eclipsing binaries together with stars whose radius has been measured by 
interferometry. Unfortunately, data are still scarce to draw any 
conclusions. This is specially true in the case of stars with 
interferometric radii since they all have very low activity levels and 
thus provide no baseline to extract any relevant trends.

An interesting consequence of the impact of activity on effective 
temperatures and radii is a possible bias in the determination of stellar 
ages from evolutionary models. This is especially relevant in the context 
of young open clusters. As discussed by Torres \& Ribas (2002) for the 
M-type eclipsing binary YY Gem, the age of active stars could appear 
systematically younger if a comparison with models was made using 
effective temperatures and radii. Such systematic effect could be as large 
as a factor of 2 for early M-type stars. However, the ages of young clusters 
are commonly determined through analysis of a colour-magnitude diagram or, 
conversely, using luminosity and effective temperature. Results indicate 
that the luminosities of active stars are not biased with respect to those 
of their inactive counterparts, but the effective temperatures are. The CMD 
diagram would thus appear shifted to cooler temperatures, an effect that 
is beautifully illustrated by the comparison of the Pleiades and Praesepe 
in the $M_{\rm v}$ vs. $(V-K)$ diagram (Stauffer et al. 2003). Young 
clusters analysed {\em only} in the region of the CMD where activity plays 
a role (i.e., G-, K-, and M-type stars) might yield ages compatible with a 
PMS phase while they are indeed already past the zero-age main sequence. 
This should not affect young cluster ages determined from the position of 
the upper main sequence turn-off since this occurs at
spectral types where stars have a radiative envelope and are therefore
magnetically inactive. However, it may have an influence on studies using
isochrones in the CMD of the lower main sequence (e.g., Jeffries et al.
2005). A simple calculation using our temperature differentials indicates
that ages of young clusters determined from (active) low-mass stars could 
be systematically underestimated by about 40\%. Interestingly, this systematic
effect agrees well with the discrepancy between CMD ages and Li depletion
boundary ages in young clusters (e.g., Barrado y Navascu\'es et al. 2004).

\section{Conclusions}

The results of the analysis fully agree with the findings from low-mass 
eclipsing binaries. Therefore, the radius and temperature discrepancies 
found in the latter are not a consequence of their mutual interactions as 
being in close binaries but can be generalised to the entire population of 
low-mass stars. By this scenario, {\em any} active star (either single or 
binary) would be equally affected by the mechanism responsible for 
lowering its photospheric temperature and increasing its radius, while 
keeping its luminosity roughly constant. In this work we thus provide 
solid ground for the interpretation of the differences between active and 
inactive stars pointed out already by Kuiper (1942), and later studied in 
more detail by Stauffer \& Hartmann (1986). The observed separation 
between the active and inactive star sequences in the H-R diagram is 
caused by systematic temperature (or spectral type) rather than luminosity 
deviations. Our conclusions stem from a comprehensive analysis using a 
large sample of stars with accurate trigonometric parallaxes, and 
therefore the results are statistically relevant.

In a recent study, Torres (2007) has determined the mass and radius of the
M2.5 transiting planet host GJ 436 from the stellar radiative properties
and constraints from the transit. The resulting radius also seems to be
larger than model predictions by about 10\% yet the star does not show any
strong activity. Note that the masses and radii determined in this way are
not fundamental (as opposed to those coming from eclipsing binaries) and that
they heavily rely on the assumed value of the effective temperature, which 
is not well established for M-type stars as discussed above. Thus, the
comparison with stellar models may not be as meaningful as that resulting
from classical eclipsing binaries. However, with the expected increase 
in the number of M stars with transiting planets, this method can provide 
valuable data to further evaluate the activity effects.

If we generalise the activity-related differences found in single and 
binary stars we conclude that current stellar structure and evolution 
models are not adequate to describe the physical (radii) and radiative 
(effective temperature, colour indices) properties of {\em all} active 
low-mass stars. The problem is particularly severe for the analysis of 
open clusters or star forming regions, where low-mass stars may display 
high-levels of magnetic activity. This is because there exists a good 
correlation between activity and age, in the sense that younger stars 
rotate faster than their older counterparts (e.g., Hartmann et al. 1984) 
and thus have more efficient mechanisms to generate magnetic energy. In 
the particular case of M-type stars (i.e., masses below $\sim$0.6 
M$_{\odot}$), which stay active up to relatively long rotation periods 
(Pizzolato et al. 2003) and, conversely, old ages (up to 1 Gyr or more), 
the discrepancy between models and observations may apply to the entire 
young disk population. We are currently working on a realistic 
implementation of the effects of magnetic activity in theoretical 
structure and evolution models to resolve the observed shortcomings. The 
appropriate framework in this direction is being established by Chabrier 
et al. (2007).

\begin{acknowledgements}
We are grateful to John Bochanski for providing us with valuable input to 
carry out this work. The referee, John Stauffer, is gratefully 
acknowledged for extremely useful and constructive suggestions. The 
authors acknowledge support from the Spanish Ministerio de Educaci\'on y 
Ciencia via grants AYA2006-15623-C02-01 and AYA2006-15623-C02-02. This 
publication makes use of data products from the Two Micron All Sky Survey, 
which is a joint project of the University of Massachusetts and the 
Infrared Processing and Analysis Center/California Institute of 
Technology, funded by the National Aeronautics and Space Administration 
and the National Science Foundation. 
\end{acknowledgements}

\end{document}